\documentclass[aps,prl,twocolumn,showpacs]{revtex4}
\usepackage{graphicx}
\usepackage{color}

\newcommand{\verdet}{{\cal V}}

\newcommand{\var}[1]{{\rm var}({#1})}
\newcommand{\PRLsection}[1]{\noindent {\it#1} -}

\makeatletter

\begin{document}

\title{Squeezed-Light Optical Magnetometry}

\author{Florian Wolfgramm, Alessandro Cer\`{e}, Federica A. Beduini, Ana Predojevi\'{c}, Marco Koschorreck, and Morgan W. Mitchell}
\affiliation{ICFO - Institut de Ciencies Fotoniques, Mediterranean
Technology Park, 08860 Castelldefels (Barcelona), Spain}

\date{25 August 2010}

\begin{abstract}
We demonstrate a light-shot-noise-limited magnetometer based on
the Faraday effect in a hot unpolarized ensemble of rubidium
atoms. By using off-resonant, polarization-squeezed probe light,
we improve the sensitivity of the magnetometer by 3.2~dB. The
technique could improve the sensitivity of the most advanced
magnetometers and quantum nondemolition measurements of atomic
spin ensembles.
\end{abstract}


\maketitle

\PRLsection{Introduction} The ability to measure magnetic fields
with high sensitivity is a key requirement in many physical,
biological and medical applications. Examples can be found in the
measurement of geomagnetic anomalies, magnetic fields in space as
well as the measurement of biomagnetic fields such as the mapping
of electric and magnetic fields produced in the brain
\cite{Stuart1964,Dougherty2006,Xia2006,Bison2009}.

Optical magnetometers, based on optical readout of magnetic atomic
ensembles, are currently the most sensitive devices. These
instruments have demonstrated sensitivities better than 1
fT/$\sqrt{\rm Hz}$, with rapid advancement in recent years
\cite{Kominis2003,Savukov2005a,Budker2007,DiDomenico2007}. Two
distinct sources of quantum noise determine the fundamental
sensitivity of this technique: the atomic projection noise and the
optical polarization noise, a manifestation of shot noise
\cite{Auzinsh2004,Kominis2008,Koschorreck2010,Shah2010}. As
today's most advanced magnetometers approach the standard quantum
noise limits \cite{Wasilewski2010} understanding these limits
becomes critical for future advances \cite{Kominis2003}.

For magnetometers based on Faraday rotation and optimized for
sensitivity, contributions from projection noise and light-shot
noise are comparable \cite{Auzinsh2004,Budker2007}, and
simultaneous reduction of both sources is advantageous. A pair of
techniques for reducing these fundamental noise sources have been
proposed, spin squeezing of the atomic ensemble
\cite{Kuzmich1997,Kuzmich1998a} and polarization squeezing of the
probe light \cite{Kupriyanov1992,Auzinsh2004}, with potential to
reduce the noise to the Heisenberg limit \cite{Shah2010}, except
in the long-time regime where spin relaxation is limiting
\cite{Auzinsh2004}. Recent experiments have demonstrated spin
squeezing using optical quantum non-demolition (QND) measurements
\cite{Windpassinger2009,Schleier-Smith2010,Leroux2010} and
application of spin squeezing in magnetometry
\cite{Wasilewski2010}. We report here reduction of the other
fundamental noise source in optical magnetometry: we demonstrate
an optical magnetometer with sensitivity better than the
shot-noise limit using a polarization-squeezed probe tuned near
the atomic resonance.

In other fields of optical science, the application of squeezed
light has already been demonstrated, such as in polarization
interferometry \cite{Grangier1987}, atomic spectroscopy
\cite{Polzik1992}, and gravitational wave detection
\cite{McKenzie2002,McKenzie2004,Goda2008}.

We note that the QND measurements used to produce spin squeezing
are performed by the same mechanism as the spin readout, and are
themselves fundamentally limited by optical shot noise
\cite{Auzinsh2004, Koschorreck2010}. In that context,
polarization-squeezed probing implies a greater degree of spin
squeezing. Ultimately, it will therefore be desirable to employ
both techniques in the same experiment \cite{Soerensen1998}.

The magnetometer consists of a source of polarization-squeezed
light, a rubidium vapor cell at room temperature and a
shot-noise-limited polarimeter. By the Faraday effect, an axial
magnetic field creates a circular birefringence in the vapor. The
resulting rotation of the polarization plane of a linearly
polarized input beam is seen in the detected signal. This rotation
is described in terms of the probe beam Stokes parameters $S_0 =
I_H+I_V, \ S_x = I_H-I_V, S_y = I_D-I_{\bar{D}}, \ S_z = I_R-I_L$,
where $I$ are the intensities of the different polarization
components ($H$: horizontal, $V$: vertical, $D$: diagonal,
$\bar{D}$: antidiagonal, $R$: right circular, $L$: left circular).
The detected signal is
\begin{equation} S_y^{({\rm out})} = S_y^{({\rm
in})} + S_x (\verdet B_z + \alpha F_z)l \ ,
\label{Eq:SyInputOutput}
\end{equation}
where $\verdet$ is the Verdet constant of the vapor, $\bf B$ is
the magnetic field, $\alpha$ is proportional to the vector
component of the atomic polarizability, $\bf F$ is the collective
atomic spin, and $l$ is the length of the medium. For a
horizontally polarized probe beam, $\left< S_x \right>$ is maximal
and $\left<\right.S_y^{({\rm in})}\left.\right>$ is zero. The
magnetometer signal comes from the terms $\verdet B_z$ and $\alpha
F_z$, the latter being sensitive to field-induced spin precession.
Projection noise is present in $F_z$, while shot noise is present
in $S_y^{({\rm in})}$. We work in a regime where these fundamental
noise sources are dominant, to show clearly the advantage of
squeezed light for optical magnetometry.

In one usual mode of operation, a magnetometer operates via
precession of a polarized spin, the initial polarization rotating
into the $z$ direction in response to the field, e.g., from $x$
toward $z$ due to $B_y$ as $\left<F_z\right> = |F| \mu_0 g B_y
\tau$, where $g$ is the Land\'{e} factor, $\mu_0$ is the Bohr
magneton, and $\tau$ is the precession time \cite{Budker2007}.
This gives a gain due to precession of $G_y \equiv {\partial
S_y^{({\rm out})}}/{\partial B_y} = S_x \alpha \mu_0 g \tau |F|
l$. Technical noise sources, e.g., in the initial orientation of
$\bf F$, and environmental noise in $\bf B$ contribute to $\var{
S_y}$ as $G_y^2$, i.e., as $|F|^2$. Similarly, $G_z \equiv
{\partial S_y^{({\rm out})}}/{\partial B_z} = S_x \verdet l$, with
associated technical noise. While important progress has been made
toward reducing technical and environmental noise below the
quantum noise \cite{Budker2007, Wasilewski2010}, this is far from
trivial and we adopt the simpler strategy of reducing the gain by
reducing $|F|$. We work with an unpolarized ensemble, i.e., a
thermal distribution within the hyperfine and Zeeman levels, with
$\left< {\bf F}\right>=0$. $G_y$, the gain due to precession and
the associated technical noise are then zero, while $G_z$ remains
and we operate in the Faraday rotation mode.

The fundamental noise sources are largely unchanged in this mode
of operation, and we can demonstrate shot-noise-limited
performance under conditions that would be present in a
highly-sensitive magnetometer with greatly reduced technical
noise. The thermal distribution has intrinsic spin noise
$\var{F_z} = F(F+1) N_A/3$, compared to $\var{F_z} = |F|/2 = F
N_A/2$ for an ideal polarized state \cite{Koschorreck2010}. In the
experiment below, the light is tuned close to the transitions from
the $F=2$ manifold, which contains $5 N_A/8$ atoms and for which
$F(F+1)/3=2$. The resulting spin noise detected via the last term
in Eq. (\ref{Eq:SyInputOutput}) is $\approx 5 N_A/4$, versus
$\approx N_A$ for a fully polarized $F=2$ ensemble. The shot-noise
contribution is unchanged. In this way, we can see the full
effects of fundamental noise sources, but with a greatly reduced
sensitivity.

\begin{figure}[b]
\centering
\includegraphics[width=0.47\textwidth]{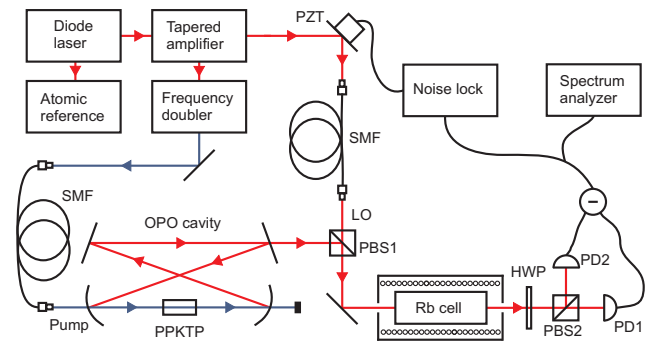}
\caption{Experimental apparatus. Rb cell, rubidium vapor cell with
magnetic coil and magnetic shielding; OPO, optical parametric
oscillator; PPKTP, phase-matched nonlinear crystal; LO, local
oscillator beam; PBS, polarizing beam splitter; HWP, half-wave
plate; SMF, single-mode fiber; PD, photodiode. \label{img:Setup}}
\end{figure}

\PRLsection{Experimental Setup} The experimental setup is shown
schematically in Fig.~\ref{img:Setup}.
As principal light source we use an external-cavity diode laser at
794.7~nm, tunable over the D$_1$ transition of atomic rubidium.
The frequency can be stabilized by FM saturated absorption
spectroscopy to individual transitions of the D$_1$ line of Rb.
The laser output passes through a tapered amplifier and is split
in two parts: The weaker part is spatially filtered with a
single-mode fiber and serves as local oscillator (LO) beam. The
stronger part is frequency doubled to 397.4~nm and then sent
through a single-mode fiber for mode-cleaning. After the fiber a
power of 42~mW is used to pump a subthreshold optical parametric
oscillator (OPO) in which squeezed vacuum is produced. The
nonlinear medium in the OPO is a type-I phase-matched PPKTP
crystal. The cavity is actively stabilized by using a
frequency-shifted beam with a polarization orthogonal to the
polarization of the squeezed vacuum. Further details of the OPO
setup can be found in \cite{Predojevic2008}.

The vertically-polarized cavity output is combined with the
horizontally-polarized LO at a polarizing beam splitter (PBS1)
with a degree of overlap of 99\%. The resulting light is
horizontally polarized, with squeezed fluctuations in the diagonal
or circular polarization basis. The polarization-squeezed light is
then sent through a 15~cm-long atomic cell at room temperature.
The isotopically purified atomic vapor contains $>$99\% $^{87}$Rb
with a small concentration of $^{85}$Rb. We lock the laser to the
5$^{2}$S$_{1/2}$(F=3)$\rightarrow$5$^{2}$P$_{1/2}$(F'=2)
transition of the D$_1$ line of $^{85}$Rb. This corresponds to a
detuning of about 700~MHz from the closest $^{87}$Rb resonance.
The cell is contained within a single-layer $\mu$-metal cylinder
to shield external magnetic fields while a coil within the
cylinder generates the desired field $B_z$.

The optical rotation is detected by a shot-noise-limited
polarimeter: after a half-wave plate at $22.5^{\circ}$, a
polarizing beam splitter (PBS2) splits the horizontally and
vertically polarized components of the beam and directs them to
the two photodiodes of a balanced amplified photo-detector with a
quantum efficiency of 95\%.
\begin{figure}[b]
\centering
\includegraphics[width=0.5\textwidth]{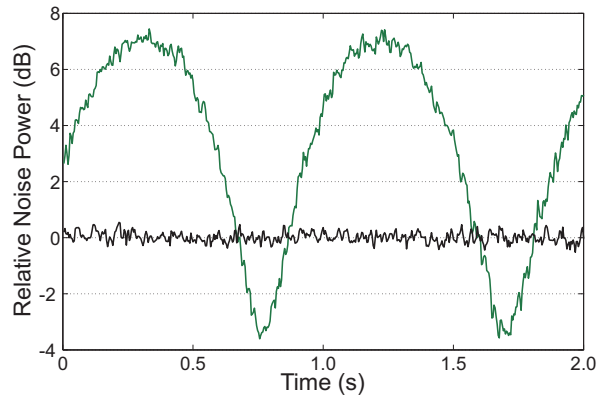}
\caption{Polarization squeezing after the atomic vapor cell.
Polarization noise power as the phase of the local oscillator is
scanned. Center frequency 1~MHz, zero-span mode, RBW=30~kHz,
VBW=30~Hz. Horizontal trace shows noise with a polarized (but not
squeezed) probe, i.e. with OPO off, and is taken as the reference
0~dB. Oscillating trace shows noise with OPO on, including regions
below the shot-noise level.\label{img:McD}}
\end{figure}
The signal is monitored on a spectrum analyzer. Quantum noise
locking is used to stabilize the phase of the local oscillator at
maximum squeezing or anti-squeezing \cite{McKenzie2005}.
\\
\\
\PRLsection{Polarization squeezing} We first characterize the
polarization squeezing at the output of the vapor cell, in the
absence of an applied magnetic field. The production of
polarization squeezing is a phase-sensitive process, with the
relative phase of the squeezed vacuum and local oscillator
determining the angle of the polarization-squeezing ellipse in the
$S_y, S_z$ plane \cite{Bowen2002}.

The polarization noise is detected with the spectrum analyzer as
the LO phase is scanned with a piezo-electric actuator, giving
rise to the squeezing trace shown in Fig.\ \ref{img:McD}. The
electronic noise is everywhere more than 13~dB below the
shot-noise level and is subtracted from data. The squeezing level
is consistent with squeezing we observed in other measurements
that were carried out without the atomic cell. The minimum of the
noise level in the squeezed phase is -3.6~dB below the shot-noise
level and the maximum 7.4~dB above shot noise in the anti-squeezed
phase. To our knowledge this is the highest degree of squeezing
obtained in a diode-laser-pumped system.

This measurement was performed at a central frequency of 1~MHz
with zero span and a resolution bandwidth of 30~kHz, a video
bandwidth of 30~Hz and a sweep time of 2~s.
\begin{figure}[b] \centering
\includegraphics[width=0.5\textwidth]{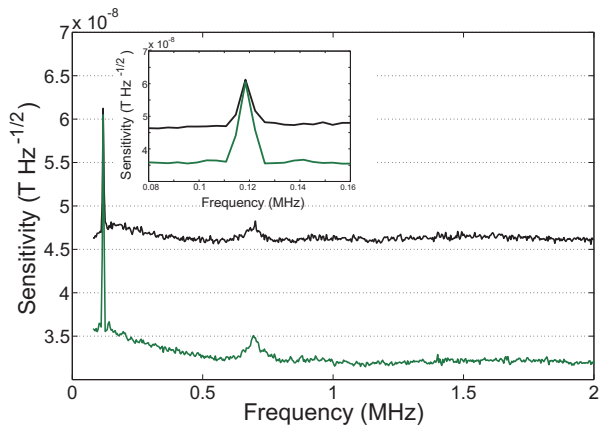}
\caption{Faraday rotation measurement. Power of the polarization
signal as center frequency is scanned, RBW=3~kHz, VBW=30~Hz. The
(upper) black curve shows the applied magnetic signal at 120~kHz
above the shot-noise background of a polarized (but not squeezed)
probe. The (lower) green line depicts the same signal with
polarization-squeezing. A zoomed view around the calibration peak
at 120~kHz is shown in the inset. \label{img:Rotation}}
\end{figure}
The total detection efficiency after creation is 82\% and includes
the escape efficiency (96\%), the homodyne efficiency (98\%),
transmission through the atomic cell (97\%) and the optical
elements (95\%), and the quantum efficiency of the detector
(95\%). The parametric gain, defined here as the ratio between the
maximum transmission of a classical beam through the cavity with
and without the presence of the co-propagating pump beam was
measured to be 4.8.

\PRLsection{Squeezing-enhanced Faraday rotation measurement} To
measure the magnetometric sensitivity, we observe the Faraday
rotation signal in response to an applied sinusoidal magnetic
field at a frequency of 120~kHz. The sensitivity is measured with
two different input polarization states: a coherent polarization
state (OPO off) and a state squeezed in $S_y$. Quantum noise
locking is used to stabilize the LO phase during the measurements.
In both cases the average polarization is horizontal, due to the
strong LO contribution, but the quantum fluctuations differ. As
shown in Fig.~\ref{img:Rotation}, the observed power spectrum in
both cases shows the reference signal due to the applied
oscillating magnetic field at 120~kHz above differing noise
backgrounds.

The LO beam has a power of 620~$\mu$W and a beam waist of 950
$\mu$m inside the vapor cell. For this intensity, beam shape, and
detuning, the magnetometer operates in a regime of nonlinear
magneto-optical rotation (NMOR) \cite{Budker2007}. A small
fraction of the atoms are optically pumped while passing through
the linearly-polarized probe beam, creating coherences within the
$F=2$ manifold. Rotation of these coherences by the $z$-polarized
magnetic field creates the conditions for alignment-to-orientation
conversion \cite{Budker2000a,Budker2002,Lombardi1969}, again by
the probe beam. Measurements of rotation angle vs input power show
a quadratic scaling consistent with this nonlinear mechanism.
Unlike optical self-rotation \cite{Fleischhauer2000a,Ries2003},
this nonlinearity does not strongly couple optical noise into
$S_y$, so long as the rotation angle remains small. The rotation
angle was calculated to be $\phi = (I_1-I_2)/(I_1+I_2)= 1.2$
$\mu$rad, where $I_{1,2}$ are the beam intensities at the two
detectors. The spectrum analyzer frequency is scanned from 80~kHz
to 2~MHz, in a sweep time of 8~s. The resolution bandwidth and the
video bandwidth were set to 3~kHz and 30~Hz, respectively and the
signal was averaged over 130 cycles.
\\
The polarimeter signal was calibrated against a linear magnetic
field sensor inserted within the coil and shielding, thus
permitting a direct conversion from measured voltage to axial
magnetic field $B_z$. The sensitivity, i.e., field noise density
as measured with the spectrum analyzer, is $4.6\cdot 10^{-8}$
T/$\sqrt{{\rm Hz}}$ for a polarized input, and reduced by 3.2 dB
to $3.2\cdot 10^{-8}$ T/$\sqrt{{\rm Hz}}$ with a
polarization-squeezed input.
It should be noted that the squeezing extends over $>2$ MHz of
bandwidth, allowing magnetic field measurements in the
$\mu$s-regime with squeezing-enhanced sensitivity. This technique is
thus also suitable to improve $\mu$s-scale QND measurements
\cite{Koschorreck2010}.

\PRLsection{Conclusions} We have demonstrated the
squeezing-enhanced measurement of a magnetic field with a hot
atomic vapor of $^{87}$Rb atoms. The measurement is
shot-noise-limited, and using a polarization-squeezed probe we
improve the sensitivity 3.2~dB beyond the shot-noise level. This
result complements recent demonstrations of spin squeezing to
reduce spin projection noise, the other fundamental noise source
in optical magnetometry. The squeezing-enhanced sensitivity
extends over a bandwidth greater than 2 MHz, allowing
high-bandwidth, sub-shot-noise magnetometry. The demonstrated
technique could be applied in advanced optical magnetometers and
in $\mu$s-scale QND measurements.

\begin{acknowledgments}
We are grateful to M. Napolitano and R. J. Sewell for helpful
discussions and to E. Polzik for drawing our attention to Ref.
\cite{Kupriyanov1992}. This work was supported by the Spanish
Ministry of Science and Innovation under the Consolider-Ingenio
2010 Project ``Quantum Optical Information Technologies'' and the
ILUMA project (No. FIS2008-01051) and by an ICFO-OCE collaborative
research program. F.~W. and A.~P. are supported by the Commission
for Universities and Research of the Department of Innovation,
Universities and Enterprises of the Catalan Government and the
European Social Fund.
\end{acknowledgments}


\end{document}